# J-aggregates of thiacyanine dye organized in LB films: effect of irradiation of light

Syed Arshad Hussain, D. Dey, S. Chakraborty, D. Bhattacharjee

Department of Physics, Tripura University, Suryamaninagar-799130, Tripura, India,

Email: sa_h153@hotmail.com

Phone: +91 9862804849 (M) Fax: +91 381 2374802 (O)

**Abstract:**

In the present communication we report the preparation and characterizations of Langmuir and Langmuir-Blodgett films of a thiacyanine dye $N,N'$-dioctadecyl thiacyanine perchlorate (NK) mixed with Octadecyl trimethyl ammonium bromide (OTAB). The relationship between the molar ratio of OTAB and NK and the orientation of molecules at the air-water interface was investigated using surface pressure – area per molecule (π–A) isotherm. UV-Vis absorption and fluorescence spectroscopic investigations reveal that prominent J-aggregation of NK molecule was observed in the LB films lifted at higher surface pressure. This J-aggregation can be controlled by diluting the NK molecules with OTAB. It was observed that the J-aggregates of NK decayed to monomer and H-aggregates when the NK-LB film was exposed to a monochromatic light of wavelength 460 nm ($\lambda_{max}$ of J – aggregates).

## 1. Introduction

The effects of molecular arrangement upon the physicochemical behaviour in molecular assemblies are current topics of research. Langmuir–Blodgett (LB) technique is an important method among different thin film preparing techniques for arranging various kinds of molecules into the form of monolayer assemblies, which may be suitable for constructing self-asembling nano sized monolayer aggregates with various functionality [1-5]. Conventionally, amphiphilic molecules in LB films are perpendicularly oriented to the LB film plane, and the thickness of LB films is integer multiples of unit molecular length. Thus, it is relatively easy to form a periodic system in direction of out-plane of LB films [2].

J – aggregates of organic dyes have attracted considerable interest as materials for spectral sensitization, optical storage and ultrafast optical switching [6]. Cyanine dyes have drawn a large attention as materials investigated for the mechanism of photosensitization [7] and for the application to photoelectric cells, non-linear optical elements and so on [8]. It is well known that these dyes form aggregates. Two-dimensional aggregates or crystallites of these dyes which are prepared by LB technique or by adsorption at an interface are investigated with special interest in their electronic and optical properties. Above all, J-aggregates [8-12] have been studied extensively which are two-dimensional crystals with chromophores arranged like a brick-stone model in a monolayer at the air-water interface [13-17].

The two-dimensional structure and order within a single monolayer as a building unit of multilayer assemblies have to be controlled carefully in order to prepare the functional superstructures for their applications [18]. The pioneering work by Kuhn and co-workers [19-20] on surface active cyanine dyes has opened the studies of J-aggregation in the two-dimensional (2D) structures formed in the LB films.

Vranken et al. [21] investigated the domains of J-aggregates formed upon adsorption of the thiacarbocyanine dye THIATS ($3,3'$-disulfopropyl-$5,5'$-dichloro-9-thylthiacarbocyanine) onto a Langmuir film of the oppositely charged amphiphile dioctadecyldimethylammonium bromide at the air-water interface. They proposed a model of radial growth of aggregated THIATS molecules, from a nucleation site into circular domains where the dye molecules adopt a brickstone arrangement. Vranken et al [22-23] also reported the influence of the deposition method and molecular structure on the topography and spectroscopy of J-Aggregates of a thiacarbocyanine dye adsorbed to a Langmuir Film. Tian et al. [24] showed that when dioctadecyl dimethylammonium bromide (DODAB) is compressed on a sub phase containing $3,3'$disulfopropyl-$5,5'$-dichlorothiacyanine (THIAMS), adsorption of the dye to the DODAB monolayer results in the formation of J-aggregates which spontaneously organize into polygonal domains of micron size. The features of the domains depend on the surface pressure. Influence of surface pressure and adsorption time on the size and shape of J aggregates of $3,3'$-disulfopropyl-$5,5'$-dichlorothiacyanine sodium salt (THIAMS) adsorbed onto dioctadecylammonium bromide (DODAB) monolayer were also investigated [25]. It was observed that J – aggregates of merocyanine dyes in LB films decayed due to irradiation of monochromatic light [26-27]. However, it is not well known what parameters are essential to control such aggregates in films fabricated by LB technique. Therefore, it is difficult to control the structure of the dye assemblies at micrometer scale. So in depth investigations of such system is of fundamental importance. Also to best of our knowledge, the effect of irradiation of monochromatic light on J – aggregates of thiacyanine dye NK in LB films have never been reported.

In our previous works we have investigated the spectroscopy of NK along with a laser dye octadecyl rhodamine B organized in LB monolayer in presence and absence of elementary clay mineral sheets [28]. Also effect of nano-clay sheet laponite on the J – aggregate of NK in LB monolayer film have been reported [29].

In this paper, we report the experimental results of surface pressure – area (π–A) per molecule isotherms of Langmuir (L) film and absorption and fluorescence spectra of monolayer LB films of a thiacyanine dye $N,N'$-dioctadecyl thiacyanine perchlorate (NK). J-aggregates of NK molecules were observed in LB monolayer deposited at higher surface pressure. It was observed that J-aggregates can be controlled by mixing OTAB in the monolayer films. It was also observed that the NK- J aggregates decayed to monomer and H- aggregates due to irradiation of monochromatic light.

## 2. Experimental

NK [Hayashibara Biochemical Laboratories Inc] and octadecyl trimethyl ammonium bromide (OTAB) [Sigma-Aldrich] were used as received. Molecular structures of NK is shown in the inset of figure 1. The main feature of the structure of the dye is characterized by a donor and an acceptor nucleus connected with a central conjugated chain. This kind of structure is favourable for the occurrence of intermolecular charge transfer, depending on the molecular environment [30].

Working solution was prepared by dissolving the NK/OTAB in HPLC grade chloroform [Acros Organics, USA]. In order to obtain Langmuir films at the air-water interface, a small amount of the dilute chloroform solution ($10^{-3}$M) of the compounds investigated were spread onto the LB trough (APEX-2000C, India) filled up with ultrapure millipore water. Allowing 15 min to evaporate the solvents, the barrier was compressed at a rate of 10 mm / min to record the surface pressure – area ($\pi$–A) per molecule isotherm. The surface pressure ($\pi$) versus average area available for one molecule (A) was measured by a Wilhelmy plate arrangement, as described elsewhere [31-32]. Stability tests for the Langmuir films were done by checking the variation of area per molecule keeping the surface pressure constant during 1 h. The films were found to be stable. Data for $\pi$–A isotherms were acquired by a computer interfaced with the LB instrument. The isotherms were reproducible within an error of 0.02 nm$^2$. Each isotherm was obtained by averaging at least five runs.

Polished smooth quartz plates were used as the solid substrates with hydrophilic surface for LB film fabrication. The substrates were dipped and raised through the floating Langmuir film vertically with a speed of 5 mm / min at a desired fixed surface pressure to prepare monolayer LB films. The transfer ratio was estimated by calculating the ratio of the actual decrease in the sub phase area to the actual area on the substrate coated by the layer. The values for transfer ratio was found to be $0.98 \pm 0.02$.

For spectroscopic measurement UV-Vis absorption spectrophotometer (Lamda-25, Perkin Elmer) and fluorescence spectrophotometer (LS-55, Perkin Elmer) were used. For absorption measurement of LB monolayer a clean quartz slide was used as reference. For fluorescence measurement the excitation wave length used was 430 nm.

For irradiation experiment, the LB films onto quartz substrate were exposed to 460 nm monochromatic light of 35 mW/cm$^2$ of exposure power from a xenon lamp of 500 W through a band-pass filter combined with color filters, J52529 and J52538 (Edmund Scientific Japan Co., Ltd.) in air at room temperature.

## 3. Results and discussions

### 3.1. Monolayer characteristics at air-water interface

Figure 1 shows the $\pi$–A isotherms of the monolayer films at the air-water interface for various NK : OTAB ratios along with pure NK and OTAB isotherm spread onto air-water interface.

Pure NK isotherm shows steep rising up to collapse pressure is reached with an initial lift-off area 0.84 nm$^2$. An inflection point at about 30 mN/m is observed. This may be an indication of reorientation or certain kind of phase change of NK molecules at the air-water interface at higher surface pressure. The mean molecular area ($A_0$) was obtained by extrapolation the isotherms to the zero surface pressure from the steep region (solid phase) of it. For NK monolayer the value for $A_0$ was 0.74 nm$^2$. Pure OTAB isotherm starts rising with an initial lift off area of 0.48 nm$^2$ and shows steep rising. In case of OTAB the value of $A_0$ is 0.20 nm$^2$, which well agrees with the reported results. The OTAB-NK mixed isotherm also shows steep rising with lift-off area and mean molecular area intermediate between the pure NK and OTAB isotherms (table -1). The inflection point at about 30 mN/m is remaining present in the mixed isotherm up to 10% of NK in the NK-OTAB mixed monolayer, although there was no inflection point for pure OTAB isotherm.

The miscibility or the phase separation of the two components of the Langmuir film can be determined on the basis of the shape of $\pi - A$ isotherms for various mole fractions, using the excess criterion and the surface phase rule [33]. Let us define the excess of the average area per molecule, $A^E$, at given surface pressure, as follows: $A^E = A_{12} - (A_1 N_1 + A_2 N_2)$ --- (1). Here, $A_{12}$ is the actual experimentally observed average molecular area in the two component film. $N_1$ and $N_2$ are the mole fractions of the components, and $A_1$ and $A_2$ are the single component areas at the same $\pi$ value.

If $A^E$ is equal to zero, the average area per molecule follows the additivity rule, $A_{12} = (A_1 N_1 + A_2 N_2)$, which means that, in the mixture, ideal mixing or complete immiscibility occurs. Deviation from zero, either positive or negative, indicates miscibility and non-ideal behaviour. The excess area per molecule for NK-OTAB mixtures in Langmuir films, at the fixed surface pressure of $\pi$ =10 mN/m is plotted in the inset of figure 1 as a function of mole fraction of the dye. A noticeable deviation from the additivity rule is observed. This is an indication of non-ideality among NK and OTAB in the binary mixture at air-water interface. In the present case, the deviation is predominantly positive, which means that repulsive interaction between the NK and OTAB predominates in the mixed films [33]. The values of $A^E \neq 0$ for the binary mixtures under investigation suggest that two components in Langmuir films are always well miscible.

To have more information about the monolayer at the air-water interface compressibility, (C), of the monolayer films was calculated according to

$$C = -\frac{1}{a_1}\frac{a_2 - a_1}{\pi_2 - \pi_1}$$

where $a_1$ and $a_2$ are the areas per molecules at surface pressures $\pi_1$ and $\pi_2$ respectively [33-34]. In the present case $\pi_1$ and $\pi_2$ are chosen as 10 and 30 mN/m respectively. It was observed (table 1) that the compressibility of pure NK monolayer is 12.5 mN$^{-1}$ whereas for pure OTAB film it is 10.3 mN$^{-1}$. NK-OTAB mixed films possesses compressibility values within this range. These low values of the compressibility suggest that the dye/OTAB monolayer films are rigid and incompressible, which agrees well with the steep rise and solid-condensed state of the isotherms (figure 1).

### 3.2. Spectroscopic characterizations

The normalized UV-Vis absorption and steady state fluorescence spectra of NK in chloroform solution are shown in figure 2. The NK solution absorbs light at 430 nm, which is attributed to the monomeric form of NK. The small high energy shoulder at 406 nm can be assigned to the $0 \rightarrow 1$ vibronic transition of the monomers and, eventually, a contribution of H-dimer absorption. It may be mentioned in this context that in order to check the contribution of H – dimer we measured the absorption spectra at high dye concentration in solution. Little increase of relative intensity of 406 nm band in comparison to 430 nm band is observed. It may be due to the contribution of H – dimer. It may be mentioned in this context that in several other work using cyanine dyes high energy shoulder of H – dimer along with monomeric band was observed [26]. On the other hand the NK fluorescence spectrum possesses a broad emission band centred at 483 nm that is assigned to the monomer emission of NK [28-29].

Figures 3a and b show the UV-Vis absorption and steady state fluorescence spectra of monolayer NK LB films lifted at different surface pressures viz. 10 and 30 mN/m respectively.

The absorption spectra of monolayer NK – LB films possess distinct band system within 400 – 475 nm regions with prominent peaks at 410, 433 and 462 nm. These three bands are assigned to be due to H-aggregate, monomer and J-aggregates [28-29, 35]. The monomer band is slightly red shifted by 3 nm with respect to the solution absorption spectrum. This small shift may be explained due to the change of microenvironment when NK molecules go from solution to solid films. At lower surface pressures the absorption is dominated by monomer. The intensities of H- and J – band are lesser and they are present mainly as shoulders. However with the increase in LB monolayer deposition pressure the extent of J-band increases. The intensity of H- and monomer band almost diminishes to weak humps with respect to the intensity of the J-band. The longer wavelength J-band becomes very much prominent and extremely sharp. These types of sharp longer wavelength bands are the characteristics of J-aggregates. The absorption peak of J-aggregates is sharp due to narrow distribution of intermolecular distance and aggregation number. Narrow distribution of intermolecular distance and aggregation number in J – aggregate make the width of the energy levels very narrow. As a result sharp and narrow absorption peak of J – aggregate is observed.

The fluorescence spectra of monolayer NK LB film deposited at lower surface pressures possess a broad band at around 450 – 500 nm regions [figure 3b]. This broad band is actually an overlapping of two band systems: the monomer band at around 483 nm and the J-band at 463 nm. For the film deposited at 15, 20 and 25 mN/m both these bands are resolvable . However, for the LB films deposited at higher surface pressure of 30 and 35 mN/m possess only the prominent sharp J-band with peak at around 463 nm. A close look at the absorption and fluorescence spectra of NK LB monolayer lifted at higher surface pressure reveals that both the absorption and fluorescence maxima of the J- band are extremely sharp and having almost overlapping peak positions with zero stokes shift. These are the characteristics of J-aggregates [6].

### 3.3. Dilution experiments of NK J-aggregates in OTAB

In the previous sections it has been observed that J-aggregates of NK molecules are formed for the LB films lifted at higher surface pressure. It would be very interesting if this aggregate formation can be controlled. There are few reports regarding the control of the size of nano-scale J-aggregates by diluting the dyes with fatty acid matrix [36]. Also J-aggregate forming dyes have been mixed with non J- forming dyes [37-39]. In few cases two J-aggregate forming dyes with different J-band peak position have also been mixed to control the extent of aggregation [37-39]. Luminescence properties of the mixed J-aggregate consisting of two kinds of thiacarbocyanine dyes or naphthothiacarbocyanine dyes having meso-substituent have been studied by Y. Yonezawa et al. [40] They also investigated the luminescence properties of the J aggregate of cyanine dyes in phospholipid langmuir- Blodgett films and exciton delocalization of the J-aggregate of oxacyanine dye and thiacyanine dye in LB films [41-42]. A.G. Vitukhnovsky et al. Investigated the fluorescence properties of mixed J – aggregates of quinocyanine dyes [43].

In the present case we checked this probability by mixing the dye NK with OTAB. Monolayer LB films of NK mixing with OTAB at different molar ratios have been prepared at 30 mN/m surface pressures. High surface pressure was chosen for film deposition because at high surface pressure the J-aggregates were prominent. The UV-Vis absorption and fluorescence spectra of monolayer LB films of NK:OTAB = 80:20, 50:50, 10:90 and 01:99 deposited at 30 mN/m surface pressures were checked. Figure 4 shows the UV-Vis absorption and fluorescence spectra of monolayer LB films of NK:OTAB = 01:99. The inset shows the dependence of absorbance of J-band formed at 462 nm as a function of mole fraction of NK in OTAB.

Interestingly it was observed that the absorbance of J-band decreases with the reduction of the concentration of the dye with respect to the surfactant matrix (inset of figure 4). The presence of J-band was observed in the film with molar ratios of the dye to the matrix down to 10:90. Beyond this concentration the J-band disappears. Whatever be the concentration of the dye is, the position of J-band peak wavelength was remained unaltered.

Thiacyanine dye NK is a cationic dye with positively charged head group. Normally NK molecules repeal each other in monolayer films. Therefore, probability of formation of J – aggregate is very less. However with increase in surface pressure through compression overcome the charge repulsion between the cationic NK molecules resulting favourable condition for J – aggregate to occur. Where, appropriate intermolecular orbital interactions for J – aggregation impose the requirement that the NK molecules be organized into a two – dimensional brick like array [6, 44]. On the other hand when OTAB, with positively charged head group, is mixed with NK, it causes an increase in charge repulsion in the NK – OTAB mixed films. Also with increasing OTAB concentration, number of NK molecules decreases in the mixed films. Therefore, with increase in charge repulsion as well as decrease in NK moieties in the NK – OTAB mixed films due to increase in OTAB concentration decreases the extent of J – aggregation of NK. As a result the intensity of J – band decreases.

In the J – aggregates the dye molecules form domains where dye molecules are organized into a two – dimensional brick like array [6, 44]. Few dye molecules in such domains coupled coherently and controls the spectroscopic property of the J – aggregates. While, diluting in OTAB, the number of NK domain decreases resulting in the decrease in peak absorbance intensity but the number of molecules coupled coherently in the remaining J – aggregate domains remains unaltered. As a result the peak position remains same. However, actual size of the aggregate is often much larger than the number of molecules coupled coherently [45-46] so from spectroscopic property it is not easy to explain about the size and nature of the aggregate domains. It requires further study of the system. Work is going on in this line in our laboratory.

It is interesting to mention in this context that S. Vaidyanathan et al. [47] reported the enhancement of contact and aggregate formation of cationic cyanine dyes by mixing oppositely charged coaggregate. Where oppositely charged coaggregate decreases the charge repulsion between the dye chromophore. On the other hand addition of similar charged coaggregate results the separation of monolayer into small domains of dye and decrease in the extent of J – aggregation [48].

It is worthwhile to mention in this context that in one of earlier work [29] we have demonstrated the control of J – aggregation of thiacyanine dyes in LB films by incorporating clay particles in the films. It was observed that J-aggregates of NK remain present in LB films lifted at lower as well as higher surface pressure in absence of clay particles laponite. However, in presence of clay particles J-aggregates are formed only in LB films lifted at higher surface pressure of 30mN/m and the J-aggregates of NK molecules are totally absent in the films lifted at surface pressures lower than 30 mN/m. In presence of clay platelet each NK molecule in the Langmuir monolayer neutralizes one negative charge on the laponite particles by an ion exchange reaction, which prevents the NK molecules to come close enough to form aggregates in NK – clay hybrid films. At higher surface pressures of 30 mN/m nano-trapping levels are formed at the film surface due to partial overlapping of clay platelets. Some NK molecules may slip and get squeezed to these nano-trapping to form favourable conditions for J-aggregates [29].

### 3.4. Irradiation effect of monochromatic light on J-aggregates

To have idea about the irradiation effect of visible light on the aggregation we have exposed the LB films in front of the 460 nm monochromatic light of exposure power 35 mW/cm$^2$ from a xenon lamp and measured the absorbance of the films before and after the irradiation. The normalized absorbance, ($A/A_0$) of J-aggregates, monomers and H-aggregates as a function of extent of exposure dose, where A and $A_0$ are the absorbance after and before irradiation, respectively for the LB films deposited at 30 mN/m surface pressure are shown in figure 5.

From the figure it is observed that the normalized absorbance of J-aggregates decreased due to light irradiation and those of monomers and H-aggregates increased. These results are indicative of two possibilities (i) J-aggregates in NK – LB film may decay to monomers and H-aggregates or (ii) NK molecules decay due to light irradiation. In order to clarify those two points, the NK LB films before and after light irradiation were dissolved in chloroform solution and the absorption spectra of the solutions were measured. It was observed that both the spectra are in good agreement and identical (figure not shown). This confirms that the NK molecules do not decay after light irradiation. Hence, the conclusion is that the J-aggregates decay to monomers and H-aggregates after irradiation of light. Similar effects have been reported for merocyanine [MC] dyes [26-27]. M. kushida et al. [26] reported that J – aggregates of MC molecules decayed to monomer and H – dimer upon irradiation of monochromatic light of suitable wavelength. T. Kawaguchi et al. [27] observed that J – aggregation of MC dyes in LB films containing $Co^{+2}$ ion dissociates upon irradiation of laser light, whereas, the same containing $Ba^{+2}$ ion do not dissociate upon irradiation of laser light.

## 4. Conclusions

In the present communication we have demonstrated the behaviour of NK molecules in Langmuir and Langmuir-Blodgett films. It was observed that at higher surface pressure NK molecules form strong J-aggregates in LB films. This J-aggregate was controlled by diluting of NK molecules with OTAB before film preparation. At least partial miscibility between the two components (OTAB and NK) in the binary mixture in the Langmuir films were observed. Interestingly it was observed that the absorbance of J-band decreases with the reduction of the concentration of the dye (NK) with respect to the surfactant (OTAB) matrix and for OTAB concentration 90% and above the J-band totally disappears. Interestingly it was observed that the J-aggregates decay to monomers and H-aggregates after irradiation of light.

**Acknowledgments**

The author SAH is grateful to DST for financial support to carry out this research work through DST Fast – Track project Ref. No. SE/FTP/PS-54/2007.

**Figure caption**

Figure 1: Surface pressure-area per molecule ($\pi - A$) isotherms for the Langmuir films at the air-water interface for different NK : OTAB ratio. Inset: (a) Molecular structure of NK.

(b) Plot of excess area per molecule for NK – OTAB mixtures in Langmuir films as a function of mole fraction at a fixed surface pressure of 10 mN/m.

Figure 2: Normalized absorption and fluorescence spectra of NK in chloroform solution

($10^{-6}$M).

Figure 3: (a) Absorption and (b) fluorescence spectra of monolayer NK – LB films lifted at different surface pressures viz. 10, 15, 20, 25, 30 and 35 mN/m. Inset shows the isotherm of NK and the corresponding surface pressures at which LB films were deposited.

Figure 4: UV-Vis absorption and fluorescence spectra of monolayer LB films of NK:OTAB = 01:99, deposited at 30 mN/m surface pressure.

Figure 5: Normalized absorbance ($A/A_0$) of J-aggregates, monomer and H-aggregates as a function of exposure extent for a NK – LB monolayer film deposited at 30 mN/m surface pressure.

Table-1: Monolayer characteristics taken from $\pi - A$ isotherms.

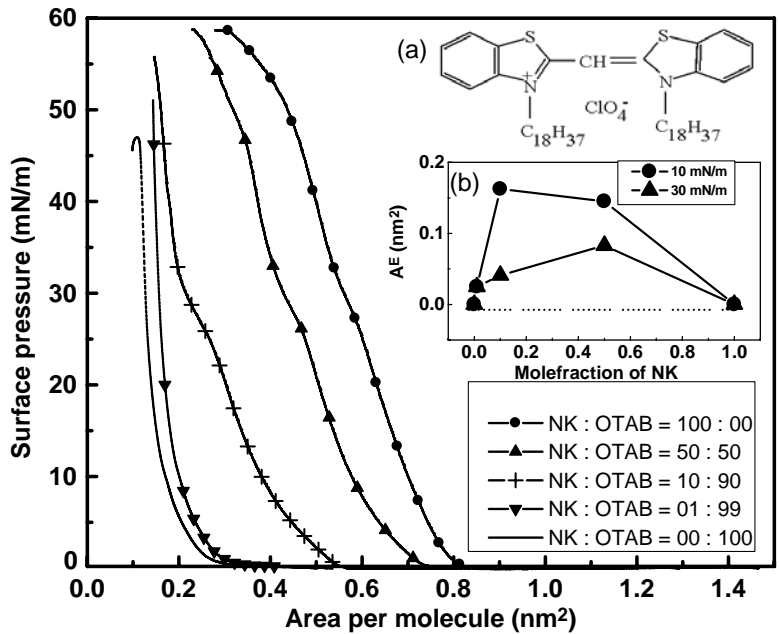

Figure 1: S. A. Hussain et. al

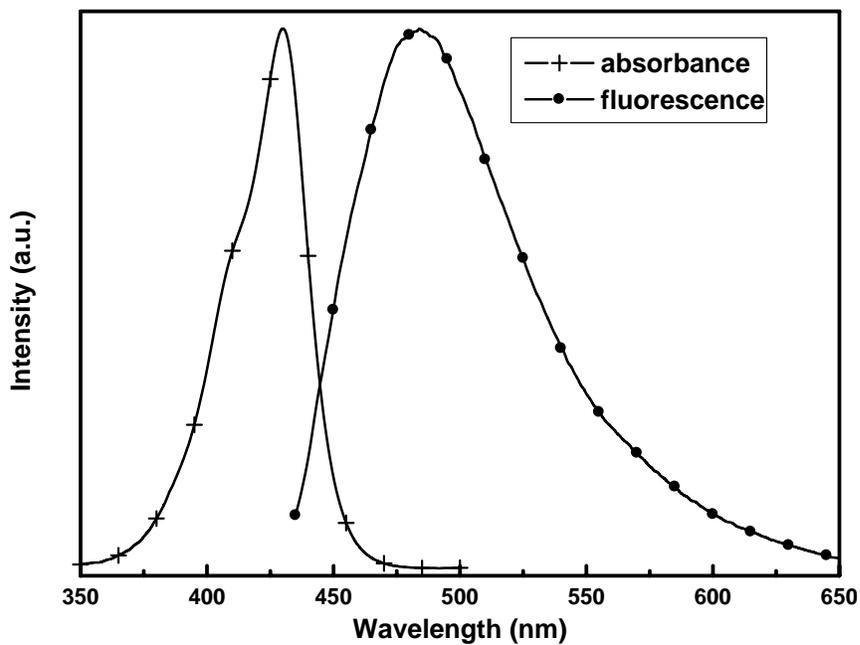

Figure 2: S. A. Hussain et. al

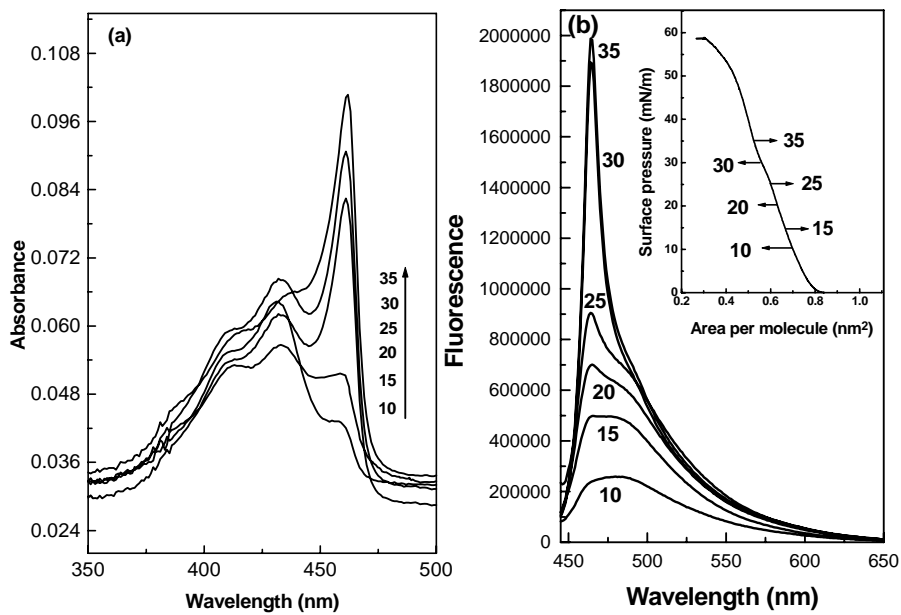

Figure 3: S. A. Hussain et. al

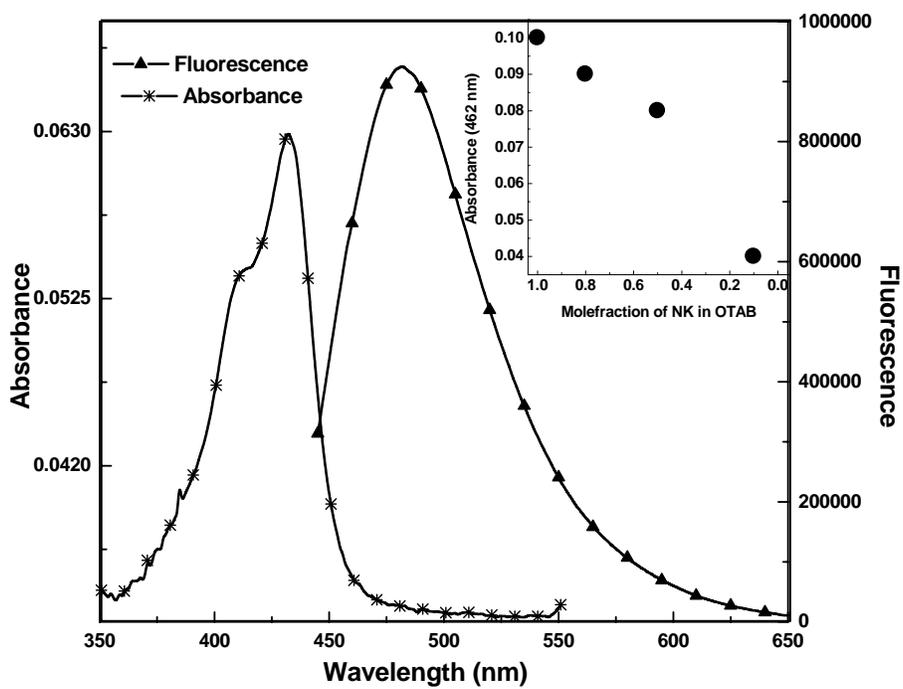

Figure 4: S. A. Hussain et. al

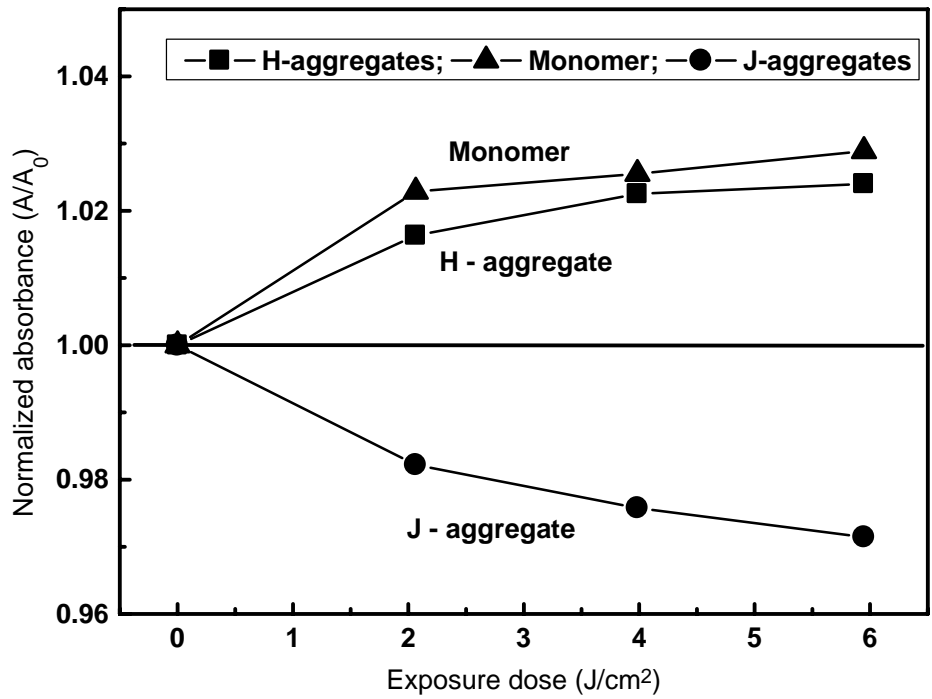

Figure 5: S. A. Hussain et. al

| NK : OTAB | Lift-off area (nm$^2$) | A$_0$ (nm$^2$) | Compressibility (mN$^{-1}$) |
|---|---|---|---|
| 100:00 | 0.84 | 0.74 | 12.5 |
| 50:50 | 0.83 | 0.57 | 12.9 |
| 10:90 | 0.59 | 0.34 | 15.8 |
| 01:99 | 0.51 | 0.22 | 9.8 |
| 00:100 | 0.48 | 0.20 | 10.3 |

Table 1: Syed Arshad Hussain et. al